\documentclass[a4paper,11pt]{article}
\usepackage{pos}

\title{Ultra-Low background germanium spectrometers at the China Jinping Underground Laboratory}

\author*[a]{Jikai Chen}
\author[a]{Zhi Zeng}
\author[a]{Hao Ma}
\author[a,b]{Jianping Cheng}

\affiliation[a]{Key Laboratory of Particle and Radiation Imaging (Ministry of Education) and Department of Engineering Physics, Tsinghua 
University,\\Beijing 100084, China}

\affiliation[b]{School of Physics and Astronomy, Beijing Normal University,\\Beijing 100875, China}

\emailAdd{cjk20@mails.tsinghua.edu.cn, mahao@tsinghua.edu.cn}

\abstract{Four ultra-low background germanium spectrometers, called GeTHU, have been installed at the first phase of China Jinping Underground Laboratory (CJPL-I), and served for material screening of dark matter and neutrino experiments. Recently, a new multi-detector spectrometer with five germanium detectors has been developed and installed at the second phase of CJPL (CJPL-II) with a minimum detectable activity (MDA) of about 10 $\mu$Bq/kg. In addition, another fifteen GeTHU-like spectrometers have been installed at CJPL-II with an MDA of about 1 mBq/kg. This paper will introduce the ultra-low background germanium spectrometers including shielding design, background characteristics and application to material screening.}

\FullConference{19th International Conference on Topics in Astroparticle and Underground Physics (TAUP2025)\\
24–30 Aug 2025\\
Xichang, China\\}

\begin{document}
\maketitle

\section{Introduction}\label{}

The rare event search experiments play an important role in the field of particle physics and astroparticle physics, including dark matter direct detection experiments and neutrinoless double-beta decay (0$\nu\beta\beta$) experiments. The materials used in rare event search experiments have low activity at the level of mBq/kg to $\mu$Bq/kg. The China Jinping Underground Laboratory (CJPL) with a vertical rock overburden of 2400 m is located in the middle of a traffic tunnel\cite{RN1}. Four ultra-low background germanium spectrometers, called GeTHU spectrometers\cite{RN2}, have been installed at CJPL-I. 

Recently, a new multi-detector spectrometer with five germanium detectors, called ARGUS, has been developed and installed at the polyethylene (PE) shielding room of CJPL-II with a minimum detectable activity (MDA) of about 100 $\mu$Bq/kg level\cite{RN3}. In addition, another fifteen GeTHU-like spectrometers have been installed at the PE shielding room with an MDA of about 1 mBq/kg level. The PE shielding room is a class 10,000 clean room with 10 mBq/m$^3$ Rn-free air supply, which has 1 m thick PE walls, effectively shielding neutrons and gamma-rays from the surrounding concrete and rock\cite{RN4}. This paper will introduce the ultra-low background germanium spectrometers, including shielding design and background decomposition.

\section{Ultra-low background spectrometers at CJPL-II}\label{}
\begin{figure*}[h]
	\centering
	\includegraphics[scale=0.32]{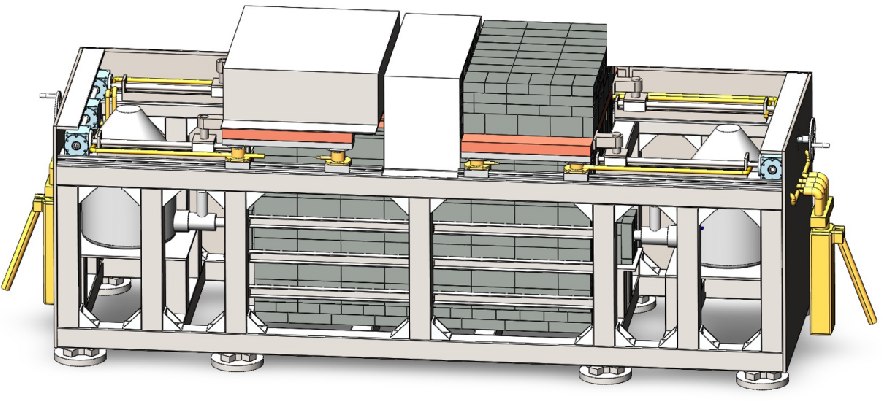}
	\caption{Detectors and shielding of the GeTHU spectrometers.}
	\label{fig:1}
\end{figure*}

The fifteen GeTHU-like spectrometers include coaxial, well, and low-energy HPGe detectors for different samples.  All detectors are proposed to have a germanium crystal with a relative efficiency of 100$\%$. From outside to inside, the detectors are surrounded by the passive shield including a 15 cm layer of ordinary lead ($^{210}$Pb<200 Bq/kg), and a 15 cm layer of high-purity copper, as shown in Fig. \ref{fig:1}. 

 \begin{figure*}[h]
	\centering
	\includegraphics[scale=0.17]{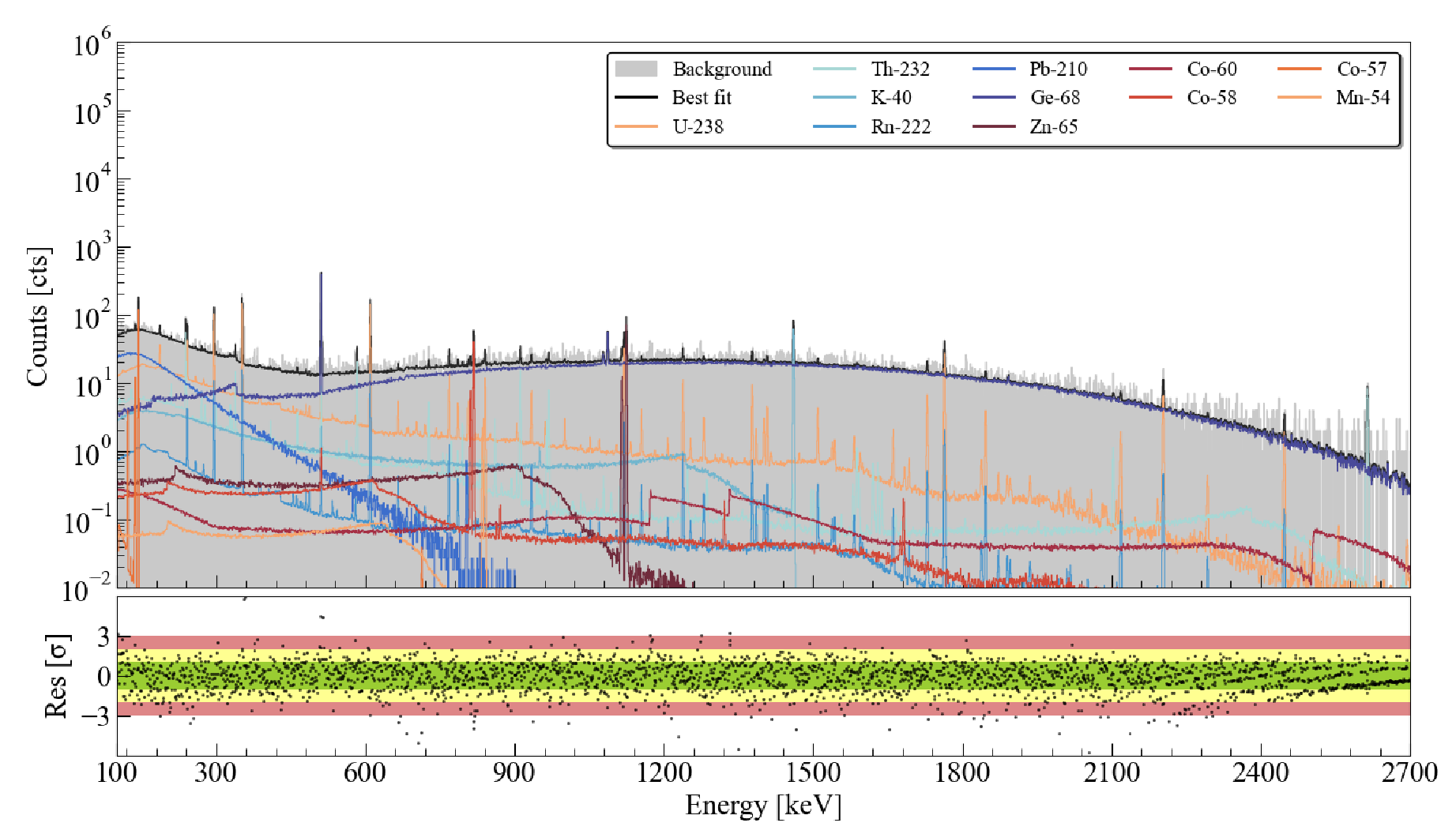}
	\caption{Background decomposition of GeTHU-\uppercase\expandafter{\romannumeral5}. The measured background spectrum is labeled in gray. The colored soiled lines are its decompositions. In the three lower panels displaying the normalized residual distributions the central 1$\sigma$-, 2$\sigma$- and 3$\sigma$-bands are marked in green, yellow, and red, respectively.}
	\label{fig:2}
\end{figure*}

The background count rates are 249.0±1.2 cpkd in the energy region of 100 keV to 2700 keV for GeTHU-\uppercase\expandafter{\romannumeral5}. The origins of the background radiation were investigated through Monte Carlo simulations and background decomposition. Statistical inference is conducted within a Bayesian framework, where the likelihood is multiplied by a factor that simulates the prior knowledge of each background component using Bayesian theorem\cite{RN5}. The background decomposition of GeTHU-\uppercase\expandafter{\romannumeral5} is shown in Fig. \ref{fig:2}, in which components referring to the same background source at different locations are aggregated.

\begin{figure*}[h]
	\centering
	\includegraphics[scale=0.38]{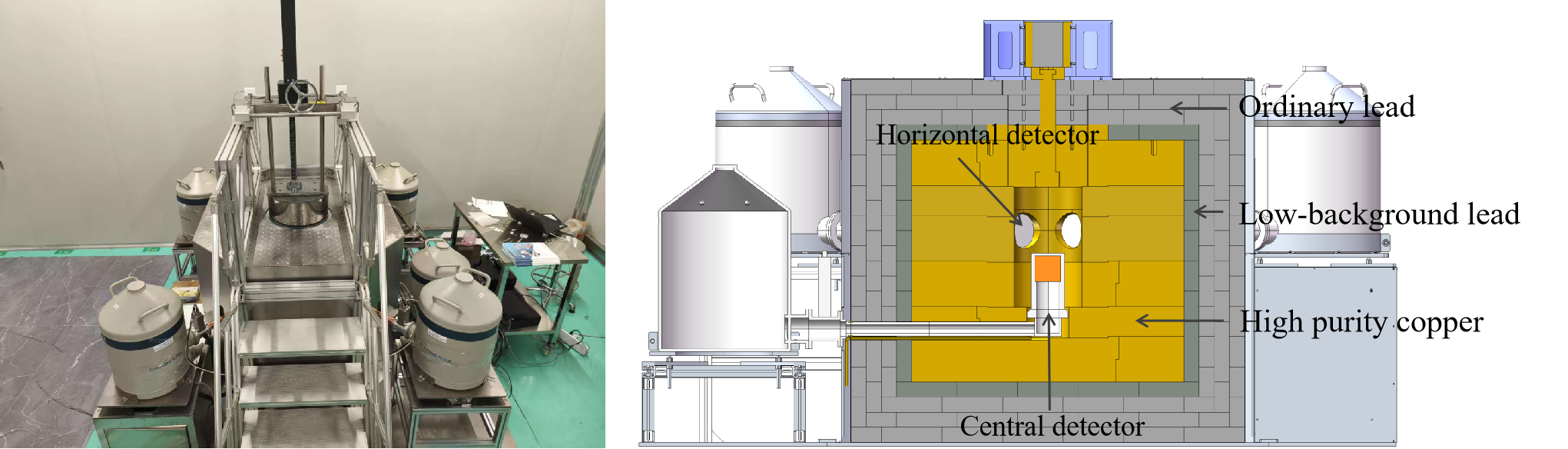}
	\caption{The layout of ARGUS spectrometer.}
	\label{fig:3}
\end{figure*}

ARGUS is a p-type coaxial germanium detector array consisting of a central detector with a U-type cold finger and four horizontal detectors with L-type cold fingers. For the central detector, the germanium crystal has a relative detection efficiency of 120$\%$. For the horizontal detectors, each crystal has a relative detection efficiency of 100$\%$. From outside to inside, the detectors are surrounded by the passive shield including a 15 cm layer of ordinary lead ($^{210}$Pb<200 Bq/kg), a 5 cm layer of low-backgroundlead ($^{210}$Pb<5 Bq/kg), and a 20 cm layer of high-purity copper, as shown in Fig. \ref{fig:3}. The sample chamber has a height of 410 mm and a diameter of 224 mm. 

 \begin{figure*}[h]
	\centering
	\includegraphics[scale=0.33]{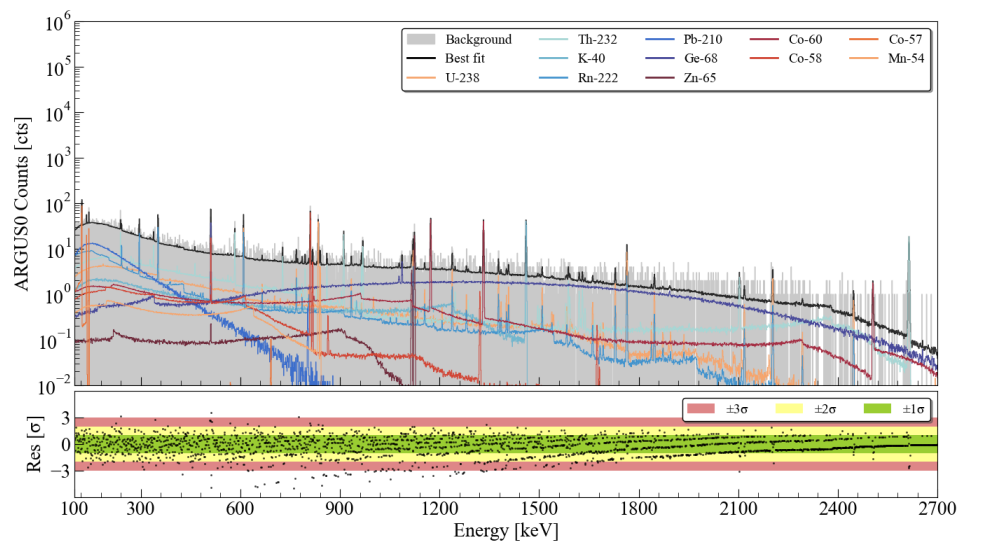}
	\caption{Background decomposition of ARGUS-D0 (central detector). The measured background spectrum is labeled in gray. The colored soiled lines are its decompositions. In the three lower panels displaying the normalized residual distributions the central 1$\sigma$-, 2$\sigma$- and 3$\sigma$-bands are marked in green, yellow, and red, respectively.}
	\label{fig:4}
\end{figure*}

The background count rates in the energy region of 100 keV to 2700 keV are 108.3±0.9 cpkd, 141.7±1.0 cpkd, 122.5±0.9 cpkd, 164.5±1.1 cpkd, and 229.2±1.3 cpkd for the five detectors respectively, based on a 60-day background measurement. The origins of the background radiation were investigated through Monte Carlo simulations and background decomposition, as shown in Fig. \ref{fig:4}. For the central detector, $^{222}$Rn contributes 13.71$\%$, cosmogenic radionuclides contribute 34.73$\%$. An copper sample (117 kg) has been screened for 42 days using ARGUS, and the MDA was 100.4 $\mu$Bq/kg for $^{214}$Bi (609 keV). 

\section{Conclusion}\label{}

Fifteen GeTHU-like spectrometers, and a new multi-detector spectrometer with five germanium detectors, called ARGUS, have been installed at CJPL-II for material screening and selection. An copper sample (117 kg) has been measured for 42 days using ARGUS, and the MDA was 100.4 $\mu$Bq/kg for $^{214}$Bi (609 keV).

\end{document}